# CORRECTION OF CROSSTALK EFFECT IN THE LOW ENERGY RHIC ELECTRON COOLER BOOSTER CAVITY

Binping Xiao[†], K. Mernick, F. Severino, K. Smith, Wencan Xu
*Brookhaven National Laboratory, Upton, New York 11973-5000, USA*

The Low Energy Relativistic Heavy Ion Collider (RHIC) electron Cooler (LEReC) is designed to deliver a 1.6 MeV to 2.6 MeV electron beam, with rms *dp/p* less than 5e-4. The superconducting radiofrequency (SRF) Booster Cavity is the major accelerating component in LEReC. It is a 0.4 cell cavity operating at 2 K, providing a maximum energy gain of 2.2 MeV. It is modified from an experimental Energy Recovery Linac (ERL) photocathode gun, and thus has fundamental power couplers (FPCs), pickup (PU) couplers (field probes) and HOM coupler close to each other on the same side of the cavity. Direct capacitive coupling between the FPC and PU, called the crosstalk effect, combined with microphonic detuning, can induce closed loop voltage fluctuations that exceed the total energy spread requirement of LEReC. The crosstalk effect in this cavity is modelled, simulated, and measured, and A correction method is proposed and demonstrated to suppress the voltage fluctuation so that energy spread requirement can be met.

DOI:

## I. Introduction

LEReC is a non-magnetized cooling approach that uses electron bunches at kinetic energies between 1.6 MeV and 2.6 MeV (to match ion beam velocities a several low energies), with rms *dp/p* less than 5e-4, to cool the ion bunches in RHIC [1]. The electron Linac of LEReC consists of a DC photoemission gun, a 704 MHz SRF Booster Cavity, and three normal conducting cavities [2, 3]. The 704 MHz SRF Booster Cavity accelerates 400 keV bunches from the DC gun near crest, with an accelerating voltage of up to 2.2 MV. This cavity was modified from an experimental ERL photocathode gun. It has two FPCs, two PUs, and one HOM damper, with another PU on the HOM damper (HOMPU) to monitor the voltage of the fundamental mode $V_{hompu}$ which leaks into the HOM damper. All couplers mentioned above are on the same side of the cavity (downstream with respect to the beam direction), shown in Figure 1. [4]

Typically, the FPC and the PU are positioned on different sides of the cavity, and thus isolated by the cavity with no direct coupling between them. In an SRF gun, however, one side of the cavity is reserved for photocathode with its stalk, thus FPC and PU are installed on the same side of the cavity that is opposite to the photocathode. With FPC and PU on the same side of the cavity, direct coupling between FPC and PU causes errors in the RF response of the PU, a so called crosstalk effect. This effect was first studied by Zhao at BNL using equivalent circuit of the cavity with couplers by applying an additional capacitor between FPC and PU [5]. Later He at Peking University proposed a simple method to extract the resonant frequency $f_0$ and loaded quality factor $Q_L$ from the RF response with crosstalk [6]. These two references mainly focused on the effect of crosstalk to the RF resonance measurements, $f_0$ and $Q_L$, etc. As pointed out by both Zhao and He, crosstalk effect severely distorts the RF response (S-parameter) between FPC and PU, $S_{21}$, when the cavity is at room temperature, corrections need to be made to retrieve the right $f_0$ and $Q_L$. While in superconducting state, the distortion of the RF response is insignificant [5].

Further study however, suggested that the crosstalk effect in the Booster Cavity can produce significant voltage errors relative to the LEReC energy spread requirement, driven by both slow frequency drift and dynamic microphonic detuning. During operation, to avoid frequent small motion and extend the lifetime of the main frequency tuner, a "dead band" is applied ot the main tuning loop. Within this "dead band", the tuner is static. While cavity's working frequency drifts, the corresponding voltage error is closed loop compensated by controlling the amplifier's amplitude and phase, without changing its frequency, to ensure a constant transmitted voltage $V_T$ on PU (both in amplitude and in phase). It is found that during operation, there is a ±1000 Hz slow frequency drift, and a ±100 Hz dynamic frequency error due to microphonics. In a cavity without crosstalk, $V_T$ is proportional to the cavity accelerating voltage $V_c$, noted as $V_c=kV_T$. In a cavity with crosstalk, since the $S_{21}$ is slightly modulated by the crosstalk effect in superconducting state, it deviates from the $S_{21}$ without crosstalk effect, which is in Lorentz distribution form. With $V_T$ held constant by the control loop, the actual $V_c$ thus deviates from $kV_T$ when the cavity resonant frequency changes.

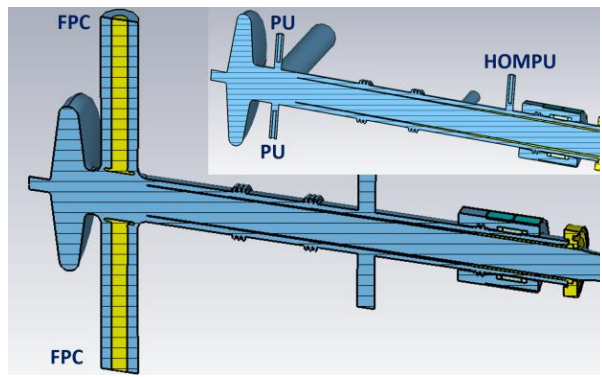

Figure 1. LEReC Booster Cavity cross-section view, with top-right plot rotated 90 degrees from the large plot. Electron beam is moving left to right. There are two FPCs and two PUs positioned symmetrically around the downstream of the beam pipe. The HOMPU is further downstream.

---

[†]binping@bnl.gov

Per LEReC specification, at 1.6 MeV, the cavity voltage error should be within ±1.12 kV peak-to-peak. Although the crosstalk effect modulation of the $S_{21}$ is quite small, the resulting voltage error could not be tolerated and need to be addressed.

## II. Circuit Model

A circuit model is used to model the crosstalk effect, shown in Figure 2. In this model, we analyse only one FPC (with admittance $B_1 = j\omega C_1$) and one PU ($B_2 = j\omega C_2$). The crosstalk is represented by a capacitive coupling between FPC and PU, $B_{12} = j\omega C_{12}$. The cavity admittance is represented by $Y_c = (1 + jQ_0 2\Delta\omega/\omega_0)/R$, with cavity shunt impedance $R = (R/Q) \times Q_0$, $R/Q$ temperature independent and unloaded quality factor $Q_0$ temperature dependent, $\omega_0$ the cavity resonant angular frequency of the working mode, $\omega$ the amplifier's driving angular frequency, and $\Delta\omega = \omega - \omega_0$ the difference between them. $V_g$ represents the voltage of the RF amplifier. $G_S \ (= 1/Z_S)$ and $G_L \ (= 1/Z_L)$ are the admittance of source and load. This circuit model is similar to the one shown in reference [5], with the difference that a term $I_b$ is added to model the cavity with electron beam, which is necessary while we apply the correction of the crosstalk. The coupling coefficients of the FPC and PU are $\beta_1 = -B_1^2 R/G_S$ and $\beta_2 = -B_2^2 R/G_L$, respectively. We further define the loaded shunt impedance $R_L = (R/Q) \times Q_L$, with loaded quality factor $Q_L = Q_0/(1 + \beta_1 + \beta_2)$. For the LEReC Booster Cavity at room temperature or 2 K temperature, the following approximations are used: $Y_c$ & $G_S$ & $G_L \gg B_1$ & $B_2 \gg B_{12}$.

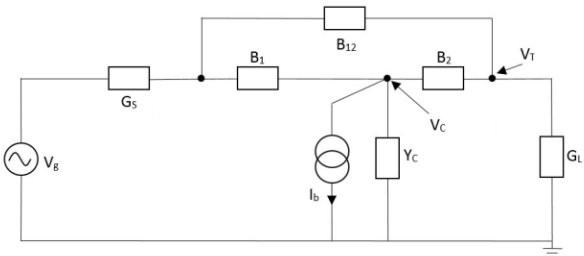

Figure 2. Circuit model of the cavity with crosstalk.

Without the crosstalk effect, $B_{12}$ does not exist, thus $V_c = kV_T$. With the crosstalk effect, $V_T$ contains voltage directly coupled from the FPC through $B_{12}$. The analysis of this circuit model should lead to the answers to the following: 1) How to measure the strength of crosstalk, 2) How the crosstalk affects cavity voltage, and 3) How to use the measured result to get the correct $V_c$ with beam. These will be discussed in the following sections.

## III. Measurement of the Crosstalk Effect

The crosstalk effect should be measured in a simple case without electron beam, $I_b = 0$ using $S_{21}$ between FPC and PU. From the circuit model we have:

$$S_{21}/2 \approx (B_{12}Y_C + B_1 B_2) / \left[ Y_C G_L + (\beta_1 + \beta_2) G_L/R \right]$$

The term that contains $B_{12}$ comes from the crosstalk effect, and the term without $B_{12}$ comes from the cavity resonance. Note both terms are $\Delta\omega$ dependent. From this expression it is not easy to separate the crosstalk term from the cavity Lorentz resonance term. This expression is further noted as:

$$S_{21}/2 \approx B_{12}/G_L + \frac{B_{12}/G_L(Q_L/Q_0 - 1) + B_1 B_2 R_L/G_L}{1 + jQ_L 2\Delta\omega/\omega_0}$$

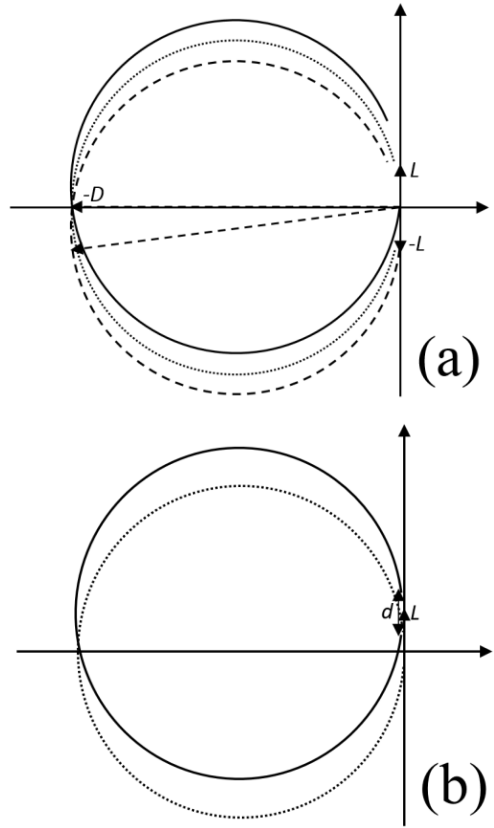

Figure 3. Polar chart of the $S_{21}$ (a) at cryogenic temperature (b) at room temperature. The dotted circle represents the cavity resonance without crosstalk, and the solid circle with crosstalk. The dashed circle in (a) represents the rotation and enlarging of the circle. This plot is for illustration only.

Figure 3 is used to help understanding this equation, with Figure 3(a) the $S_{21}$ at cryogenic temperature and Figure 3(b) that at room temperature. In Figure 3, the $S_{21}$ is shown as a solid circle. The 1st term ($B_{12}/G_L$) of this equation is a vector independent of $\Delta\omega$, in polar (or Re-Im) chart it is a translation originated from crosstalk (vector $L$ in +y axis in Figure 3). In the 2nd term, the 2nd portion

$B_1B_2R_L/G_L/(1+jQ_L 2\Delta\omega/\omega_0)$, which does not contain $B_{12}$, represents the cavity resonance without crosstalk effect, shown as dotted circles in Figure 3, and the 1st portion $B_{12}/G_L(Q_L/Q_0-1)/(1+jQ_L 2\Delta\omega/\omega_0)$, which contains $B_{12}$, represents a rotation and enlarging around the origin point due to crosstalk. At cryogenic temperature, this term (shown as the vector -$L$ in -$y$ axis in Figure 3(a)) is small, but non-zero compared with the diameter of the circle (shown as the vector -$D$ in -$x$ axis in Figure 3(a)). Note that for the Booster Cavity at cryogenic temperature, $Q_0 \gg Q_L$, thus this vector (in -$y$ axis) has the same amplitude as the vector $L$ in +$y$ axis. At room temperature, both couplers are weakly coupled to the cavity, $\beta_1$ and $\beta_2$ are both much less than 1, $Q_L = Q_0$, which means this portion is 0. The 2nd term in this expression is a Lorentz distribution, and in the polar chart it is a circle with origin corresponding to $\Delta\omega \to \pm\infty$, shown as dotted circle in Figure 3(a), and dashed circle in Figure 3(b). We use coefficient $K$ to represent the ratio between the amplitude of $L$ and $D$. Note some vectors are exaggerated in Figure 3 for illustration only.

Network analyzer measurements were done both at room temperature and at 2K to retrieve the $K$ value. The $S_{21}$ from one FPC to one PU, with the other ports properly terminated, was first measured to get the resonant frequency $f_0$. Then the complex $S_{21}$ between $f_0 - \Delta f$ and $f_0 + \Delta f$ was measured, with the distance between point $f_0 - \Delta f$ and $f_0 + \Delta f$ in the polar chart to be $d$, and the center point of $d$ to the original point of polar chart to be $L$, shown in Figure 3(b). The selection of $\Delta f$ should be big enough so that the in polar chart, $L \gg d$. During this measurement small IF bandwidth should be used to resolve small signals.

This measurement was first done at room temperature. The measured Re-Im chart is shown in Figure 4(a). Distortions appeared at the frequencies away from resonance. This is because of the coaxial/waveguide cables that cannot be excluded from this plot by cable calibration, e.g., cables inside the cryomodule. The phase changes in these cables are frequency dependent, shown in Figure 4(c). This phase change information can be retrieved from the $S_{11}$ and $S_{22}$ parameters. After taking this into account, a corrected Re-Im chart is shown in Figure 4(b). At room temperature, $Q_L = Q_0 = 5600$, $K = B_{12}/(B_1 B_2 R_{L-RT})$. $K$ is measured to be 0.18 for the LEReC Booster Cavity.

At 2K, $\beta_1 = Q_0/Q_L \gg 1$, We have

$$K = \frac{B_{12}}{B_{12}(1/\beta_1 - 1) + B_1 B_2 R_{L-2K}} \approx \frac{B_{12}}{-B_{12} + B_1 B_2 R_{L-2K}}$$

with FPC $Q_{ext} = 1.5 \times 10^5$, $R_{L-2K}/R_{L-RT} = 26.8$.

In this case the $K$ value at 2K is predicted to be 0.0067 based on the room temperature measurement, and it is measured to be 0.0069.

For both temperatures, the phase of $B_{12}/(B_1 B_2 R_L)$ is always -$\pi/2$.

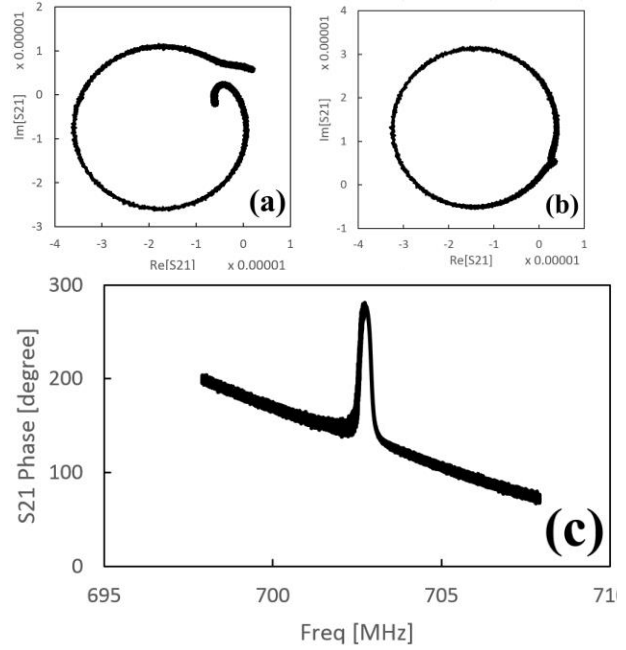

Figure 4. (a) Measured Re-Im chart of the $S_{21}$ at room temperature; (b) Re-Im chart after considering phase on coaxial/waveguide cables; (c) Measured $S_{21}$ that shows the phase versus frequency.

## IV. Crosstalk Effect and Cavity Accelerating Voltage

During the cavity commissioning in RHIC run 2018, it was noticed that $V_{hompu}$ fluctuated by a factor of 40%, while the $V_T$ was closed loop stabilized by low level RF (LLRF). The HOMPU was designed to monitor for possible multipacting in the coaxial structure of the HOM damper, see details in reference [4]. This fluctuation coincides with the LLRF tuning loop phase error that represents the cavity resonant frequency drift. The crosstalk effect between FPC and PU, and the same effect between FPC and HOMPU, are different in strength, and this difference brought the fluctuation on $V_{hompu}$. This analysis indicated that the fluctuation of $V_{hompu}$ is not caused by the multipacting effect and can be tolerated. In this case the restriction of the $V_{hompu}$ fluctuation permitted for machine protection is relaxed so that the cavity will not be falsely tripped. This phenomenon, however, brought our attention to the crosstalk effect that could cause actual cavity voltage fluctuation under closed loop operation.

This effect is first analysed using the circuit model. The difference between the solid circle and dotted circle in Figure 3(b) provides the information related to the accelerating voltage fluctuation. To separate the crosstalk effect from the Lorentz curve, one needs to translate the $S_{21}$ in Figure 3 from solid circle to dotted circle. We use the measurement at 2 K that was mentioned in the previous section. The result in amplitude versus frequency is shown in Figure 5. Based on Figure 5(b), within ±3 kHz deviation from $f_0$, the amplitude difference between the curve with crosstalk and the one without is proportional to the frequency deviation, it is 0.027dB/kHz, or -0.31%/kHz.

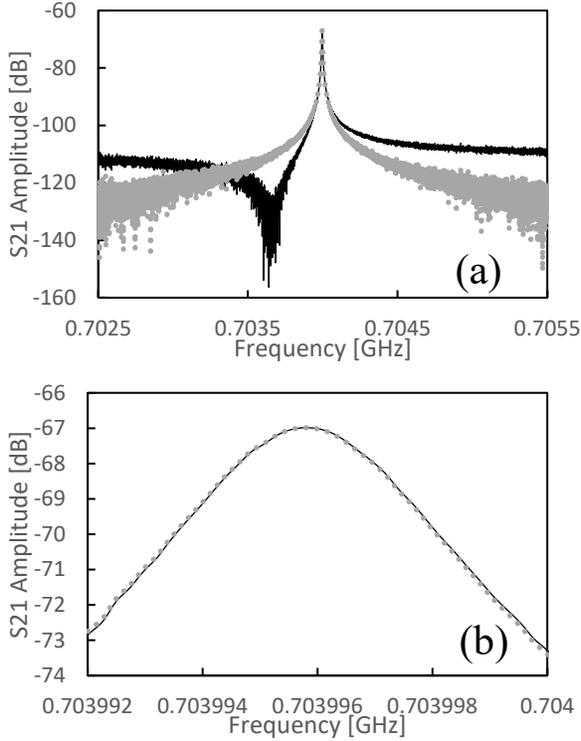

Figure 5. Correction of $S_{21}$ in dB amplitude to separate the crosstalk effect from Lorentz distribution curve, with black solid curves with crosstalk that measured at 2K, and grey dotted curves after correction. (b) is a zoom-in of (a) near the resonance.

The above results were cross-checked using the model shown in Figure 1 to numerically simulate this issue via CST [7]. In the RF simulation, two FPC ports are connected to form a new 25 Ω FPC port (port 1), and one of the PU is used as port 2. Field monitors are assigned to the on-resonant frequency and ±1 kHz, ±2 kHz, ±3 kHz off-resonant frequencies to monitor $V_c$. Note in the simulation that the power source's frequency is changing, and cavity's resonant frequency is fixed, while during operation, amplifier's frequency is fixed, and cavity's resonant frequency drifts. This simulation gives the same result as above. For comparison, $V_c$ changes with frequency drift for the case without crosstalk by -1.5×10$^{-4}$/kHz, by using a reference PU that is on the upstream side of the cavity, to suppress the crosstalk effect, which unfortunately, does not exist in the real cavity. This result implies that the simulation accuracy is not worse than -1.5×10$^{-4}$/kHz.

Based on the above calculations, with a ±1 kHz slow frequency drift, $V_c$ deviates by ±0.31%. Even if this slow frequency drift can be suppressed and we consider only a ±100 Hz frequency drift due to microphonics, the $V_c$ still deviates ±0.03%, this single effect contributed one half of the *dp/p* design budget, which was experimentally observed in LEReC operation, and thus needs to be corrected.

An additional simulation was done by cutting the PU probe by 5mm. Simulation results showed that cutting the PU probe shorter not only changes the crosstalk strength $B_{12}$ and the PU coupling strength $B_2$, and it changes the Lorentz term and the crosstalk term simultaneously, in a proportional way. Thus, making the PU probe shorter does not help.

During operation, two FPCs are driven by two LLRF systems and two amplifiers. Efforts are made trying to keep the amplitude and the phase of the signals from two amplifiers the same by controlling the LLRF, however they are still slightly different, estimated to be within 5% in the amplitude, and within 10 degrees in the phase. The S-parameter of the Booster Cavity with 2 FPCs, 1 PU and 1 reference PU on the upstream side of the cavity (without crosstalk), are modeled. With this set of S-parameter, signals with different amplitude and phase can be assigned to 2 FPCs. The power coming out of the reference PU port without crosstalk is used to monitor the $V_c$. Results showed that the crosstalk effect is not sensitive to the slight phase (10 degree) and amplitude (5%) input differences between 2 FPCs.

## V. Correction of Crosstalk Effect

Based on the analysis in the above section, in closed loop operation, the accelerating voltage fluctuation is proportional to the cavity tuning phase error, for frequency deviation within ±3kHz. This conclusion, however, cannot be used to correct the crosstalk effect since it did not include the effect of the beam current. With beam current $I_b$, the cavity accelerating voltage can be calculated based on the measured forward voltage $V_F$, reflected voltage $V_R$ and transmitted voltage $V_T$:

$$V_C = (Y_2 / B_2)V_T - (B_{12} / B_2)(V_F + V_R)$$

The first term $V_{C\_old} = (Y_2 / B_2)V_T$ is the term we normally use (without crosstalk). It does not represent the actual accelerating voltage of the Booster Cavity. The second term is an additional term, it represents the RF leakage from the tip of the FPC to the PU. The value of $B_{12}/B_2$ can be evaluated from the K value that we measured previously. Beam current $I_b$ did not appear in this equation, since it could be determined from $V_F$, $V_R$ and $V_T$.

Figure 6 shows the LLRF we developed to correct the crosstalk effect. It determines the amount of crosstalk in real time and subtracts this from the cavity PU. The LLRF feedback loops then work using the corrected PU signal to regulate the actual accelerating voltage seen by the beam.

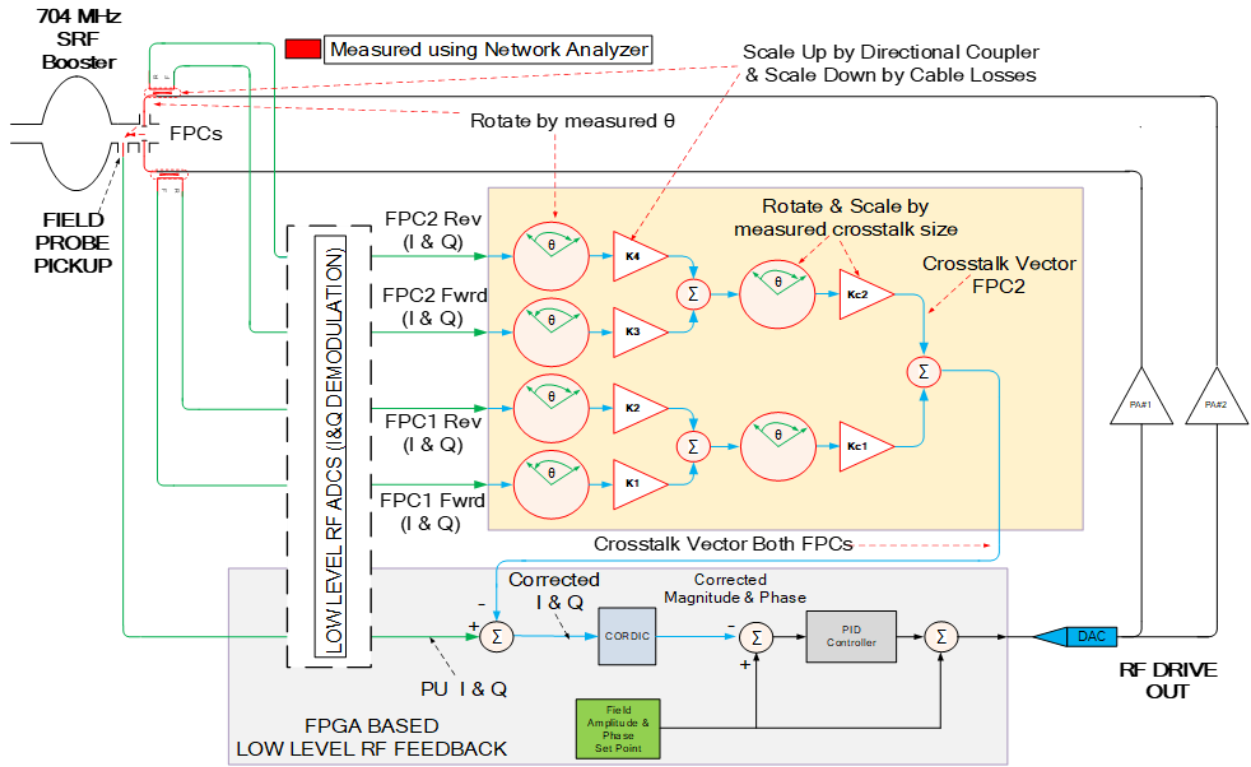

Figure 6. LLRF to compensate the crosstalk effect.

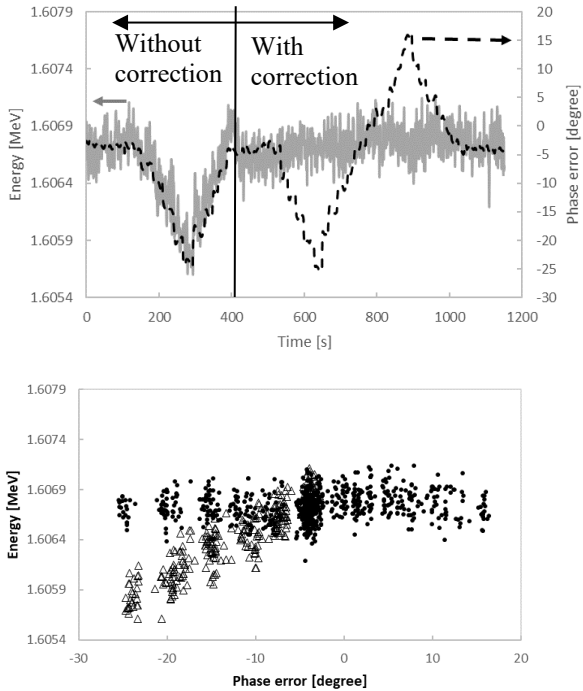

Figure 7. (top) Measured beam energy (grey solid line) without (<400s) and with (>400s) crosstalk correction, while intentionally detuning the cavity resonance frequency (black dash line), with grey solid line the beam energy, and black dotted line the phase error. (bottom) Energy versus phase error without (hollow triangle) and with (solid dot) crosstalk correction.

The crosstalk correction signals are generated in a processor on the FPGA used for the LLRF regulation of the cavity. The correction values are calculated at a 100 kHz rate. The cause for changes in FPC power are very slow drifts in cavity tuning (due to the mechanical tuner dead band) and faster microphonic detuning, which has a bandwidth less than 1 kHz. Thus, the 100 kHz update rate for the crosstalk correction is more than sufficient to correct for all sources of detuning. Please note this correction is not intended for beam loading compensation, which requires a much higher update rate that cannot be realized in this setup.

## VI. Verification of Correction Method

There is a 180-degree bending magnet in the LEReC electron Linac, so that both yellow and blue ion beams in RHIC can be cooled using the same electron beam. This bending magnet provides a way to very accurately measure the electron beam energy [8]. Without applying crosstalk correction, the cavity resonant frequency was intentionally detuned using the cavity tuner (black dash line in Figure 7), the electron beam energy was measured (grey solid line), shown in the top of Figure 7 with time <400s. The energy versus phase error is also plotted, shown as hollow triangle in the bottom of Figure 7. After applying the crosstalk correction, the cavity resonant frequency was detuned (black dash line), and the electron beam energy was measured (grey solid line), shown in the top of Figure 7 with time >400s. The energy versus phase error is also plotted, shown as solid dot in the bottom of Figure 7. From the top plot of Figure 7, one can clearly see that the electron beam energy varies with cavity detuning phase error without crosstalk correction, and after applying the correction, it does not change. Linear fitting on the bottom plot of Figure 7 suggests that without correction, energy change is about 46.5 eV per degree of detuning, and with correction, it changes to ~2.5 eV per degree of detuning. This is a reduction by a factor of about 20. Note that this is the average beam energy after the various RF gymnastics, the energy change of the beam immediately after the Booster Cavity is about 2-3 times larger.

## VII. Conclusions

In this paper, a phenomenon called crosstalk effect brought our attention in RHIC RUN 2018. Crosstalk is a direct coupling between FPC and PU that causes RF power leakage between them. While combining with the microphonics, cryogenic temperature/pressure drift, and ±20 degree "dead-band" on the phase error to reduce the movement of the frequency tuner, the crosstalk effect is found to cause cavity accelerating voltage fluctuation with a magnitude equivalent to the LEReC longitudinal energy spread budget if not properly corrected. Numerical simulation shows that this fluctuation can be explained by the distortion of $S_{21}$ simulated while there is no electron beam. To correct it, we started from a circuit model based on reference [5]. Based on the circuit model without beam, it was proposed to measure the strength of crosstalk using the $S_{21}$ of the cavity at room temperature without beam during 2018 RHIC shut down. The strength of crosstalk at cryogenic temperature was first predicted based on the room temperature measurement results, it was then measured at 2 K at the beginning of 2019 RHIC run, the measurement agrees with the prediction. Corrections were made during the 2019 LEReC commissioning, based on the analysis of the circuit model considering the case while electron beam presents. In this case the cavity accelerating voltage can be determined by an equation that is related to the measured $V_F$, $V_R$ and $V_T$, combining with the simulated $R/Q$, measured PU $Q_{ext}$, and measured strength of crosstalk. For comparison, without crosstalk, one needs measured $V_T$, combining with the simulated $R/Q$, measured PU $Q_{ext}$, to determine cavity accelerating voltage. This method is applied into the LLRF of Booster Cavity and is verified using a bending magnet to measure the electron beam energy. The energy fluctuation caused by this effect was measured to be 1kV, about 50% of the specification. It can be out of specification while combining with other effects that could enhance the energy spread like photocathode laser modulation [9], cavity HOM [2, 4], as well as the amplitude error of the control system. After correction the energy fluctuation reduced to less than 3% from this effect, which ensured the quality of the electron beam. This is an important step towards successful experimental demonstration of using LEReC to cool the RHIC gold ions [10].

## ACKNOWLEDGEMENT

The work is supported by Brookhaven Science Associates, LLC under Contract No. DE-AC02-98CH10886 with the U.S. Department of Energy (DOE). The authors would like to thank M. Blaskiewicz, J. M. Brennan, Feisi He, Tianmu Xin and A. Zaltsman for the useful discussion on the modelling and simulation, S. Seberg and R. Anderson for setting up the low power measurements, and T. Hayes, G. Narayan, S. Seletskiy, V. Schoefer, A. Fedotov for the beam measurements.